# Virufy: Global Applicability of Crowdsourced and Clinical Datasets for AI Detection of COVID-19 from Cough

Gunvant Chaudhari, Xinyi Jiang, Ahmed Fakhry, Asriel Han, Jaclyn Xiao, Sabrina Shen, Amil Khanzada[1]

*Abstract*—Rapid and affordable methods of testing for COVID-19 infections are essential to reduce infection rates and prevent medical facilities from becoming overwhelmed. Current approaches of detecting COVID-19 require in-person testing with expensive kits that are not always easily accessible. This study demonstrates that crowdsourced cough audio samples recorded and acquired on smartphones from around the world can be used to develop an AI-based method that accurately predicts COVID-19 infection with an ROC-AUC of 77.1% (75.2%-78.3%). Furthermore, we show that our method is able to generalize to crowdsourced audio samples from Latin America and clinical samples from South Asia, without further training using the specific samples from those regions. As more crowdsourced data is collected, further development can be implemented using various respiratory audio samples to create a cough analysis-based machine learning (ML) solution for COVID-19 detection that can likely generalize globally to all demographic groups in both clinical and non-clinical settings.

Impact Statement — With the spread of COVID-19, more than 73M COVID-19 cases have been found across the world. At the same time, the clinical diagnosis of COVID-19 can be time exhaustive and financially expensive for people, especially those in distant areas where COVID-19 clinical resources are limited. The ML model we introduce in this paper attempts to mediate this issue by presenting a system for detecting COVID-19 infection using cough audio recordings. Our model shows that it can be generalized to audio samples from other sources with the current training datasets. With an ROC-AUC of 77.1%, the technology can be utilized by people from different regions. Specifically, the Virufy model could offer an initial screening process for COVID-19 that individuals can access directly from their smartphones. This model would provide a cheap and free suggestion for those who do not have the resources necessary for COVID testing.

[1]The code and data used in this paper can be found: https://github.com/virufy/covid.

Gunvant Chaudhari is with the Virufy AI Research Group and University of California San Francisco, School of Medicine.
Xinyi Jiang is with the Virufy AI Research Group.
Ahmed Fakhry is with the Virufy AI Research Group and University of Alexandria.
Asriel Han is with the Virufy AI Research Group and Stanford University.
Jaclyn Xiao is with the Virufy AI Research Group and Duke University.
Sabrina Shen is with the Virufy AI Research Group and Harvey Mudd University.
Amil Khanzada is with the Virufy AI Research Group and Stanford University (e-mail: amil@virufy.ai).

## I. INTRODUCTION

As of December 16th, 2020, there were more than 73M cases of COVID-19 worldwide [1]. Widespread testing and isolation of individuals infected with COVID-19 is necessary to control infection rates and optimize healthcare resources [2]. The current gold standard of Reverse Transcription Polymerase Chain Reaction (RT-PCR) testing requires person to person contact to administer, has variable turnaround time, is expensive, and not easily accessible for the global population[3]. With cases still increasing and vaccine approval and distribution still on the horizon, accessible and affordable testing is critical to limit the pandemic.

Artificial Intelligence (AI) algorithms can be a powerful tool for a preliminary indication of a person's COVID-19 status and have been developed to accurately predict COVID-19 infection from smartphone-acquired cough sounds. With smartphone usage high and continually rising in developing countries, these devices are an ideal platform for a widespread low-cost collection of respiratory audio recordings and for implementing audio-based COVID-19 testing.

A variety of COVID-19 cough recording datasets have been collected by various groups and used to train machine learning models for COVID-19 detection. However, each of these models has been trained on data of a variety of formats and recording settings. While some, such as Coswara, collect additional counting and vowel recordings [9], others gather cough recordings exclusively [10]. Furthermore, these datasets come from various sources, such as clinical setting recorded data [10], crowdsourcing [9], and extraction from public media interviews [11]. Recordings utilize different compression formats and so far, no formal standard for gathering COVID-19 cough data has yet been defined. Variations of sources, contents and formats of datasets present a significant challenge for the ML model developers.

Research groups have explored the prediction of COVID-19 status from cough sounds, but there is still doubt about the generalizability of such models across various environments and collection methods. The largest obstacle to implementation of machine learning algorithms is their reliability on unseen data [8]. Therefore, a globally deployable ML solution needs to be trained and tested on global data to ensure high performance in all locations and situations. We present the first study showing generalizability of a multimodal deep learning model in detecting COVID-19 status on crowdsourced and clinical cough datasets from around the world, using five distinct datasets from multiple sources. As shown in Fig. 1 below, our approach consists of collecting coughs and demographics information using smartphones, extracting features, and using our ensemble

ML-model to predict COVID-19 status.

Our group, Virufy, is uniting the world to build a global artificial intelligence database of crowdsourced cough sounds to identify patterns that signify respiratory diseases such as COVID-19. The authors of this paper are current students and alumni from various institutions who have come together to tackle the pandemic.

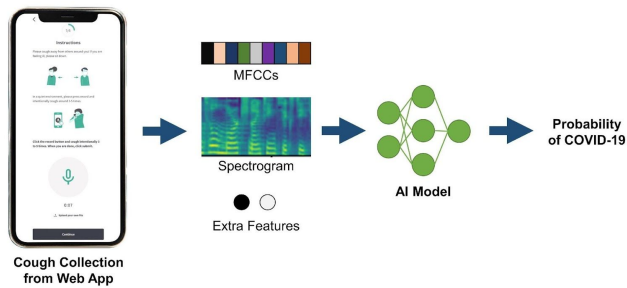

Fig. 1. Overview of our proposed cough-based COVID-19 detection system. After recordings are collected from the mobile app, the related audio features are extracted and fed into our ML model to generate probabilities of whether the person recording cough/speech is infected with COVID-19 or not.

## II. BACKGROUND

The clinical presentation of COVID-19 can be highly variable, but multiple common symptoms of the illness impact the human airway and lungs [4, 5]. Dry cough is a distinctive feature of most cases along with more severe symptoms such as pneumonia and ground-glass opacities [5]. Because of the virus's effects on the respiratory system, COVID-19 has been theorized to create unique audio signatures that are distinct from those associated with other respiratory infections. Thus, sounds such as coughing could be analyzed to detect the illness. Although these differences are difficult or impossible to detect with the human ear, ML algorithms have already shown promise in classifying cough sounds to identify respiratory diseases including pertussis, asthma, and pneumonia by using various audio features [6,7].

Various groups across the world have been focused on collecting copious high-quality data to train cough-based machine learning models, including Cambridge University in England, Carnegie Mellon University and the Massachusetts Institute of Technology (MIT) in the US, and Afeka College of Engineering in Israel. Others, such as Coswara, are focused on gathering large datasets of cough and various other human audio sounds [9]. In addition to our open dataset, Coughvid and Coswara have also open-sourced their datasets [9,10].

Work done by Cambridge University [12] and MIT [13] shows the potential of ML to detect a COVID-19 infection status. Cambridge uses mel-frequency cepstral coefficients (MFCC) and other audio statistics, such as duration, onset and zero-crossing and VGG features such as starting features [13]. The features are then further processed by Principal Component Analysis and used as input to a simple binary classifier [13]. The researchers are able to achieve an AUC of 0.82 but their dataset only has a size of 86 coughs [13]. Another approach includes using mel-Spectrogram as input to the model. A pre-trained deep convolutional neural network, ResNet-18 [16], is used and ensembles of both shallow and deep networks are explored [14]. The model is also pretrained on a larger dataset using cough/non-cough labels and techniques, such as noise augmentation, audio segmentation and time and frequency masking, are applied [14]. Both models have shown promising preliminary results in the COVID-19 classification task on Cambridge's own dataset [13] and, independently, in the Coswara dataset [9]. MIT researchers, who created a model with significant performance improvement, utilized a biomarker layer with ResNet-50 [16] based models [14]. Their best performing model inputs mel-frequency cepstral coefficients (MFCC) and uses transfer learning to output three separate biomarkers to be used as inputs to the three parallel ResNet-50 convolutional neural networks [14].

However, research published to date shows results restricted to datasets which are often not open-source, so it is difficult to judge the ability of these models to identify COVID-19 using arbitrary cough recordings. Especially in the case of using crowdsourced data as testing data where the audio samples are recorded from various environments usually with background noise, models trained on clinical data have a high possibility of not being able to generalize. In addition, prior research highlights the importance of using transfer learning and pretraining to enlarge the training data size, because the size of a single dataset may not be big enough to train deep neural networks. In this paper, we show that our model can adapt to enlarged training and testing datasets from various sources.

## III. METHODS

### A. Crowdsourced COVID-19 Training Data

We trained a deep neural network using the openly available Coswara (n=1,543) and Coughvid (n=20,072) datasets of cough sounds, with COVID status labels. From the Coswara datasets, samples with 'covid-status' of 'positive_mild', 'positive_moderate', and 'positive_asymp' were classified in the positive class (n=1,334), and all other statuses were classified as negative (n=98). Only shallow cough files were used. All were uncompressed audio files.

From the Coughvid dataset, samples with the 'COVID-19 Positive' label were classified in the positive class (n=441) and 1,000 other random samples were selected to be in the negative class to maintain data balance. The distribution of this data is shown in Fig. 2 below.

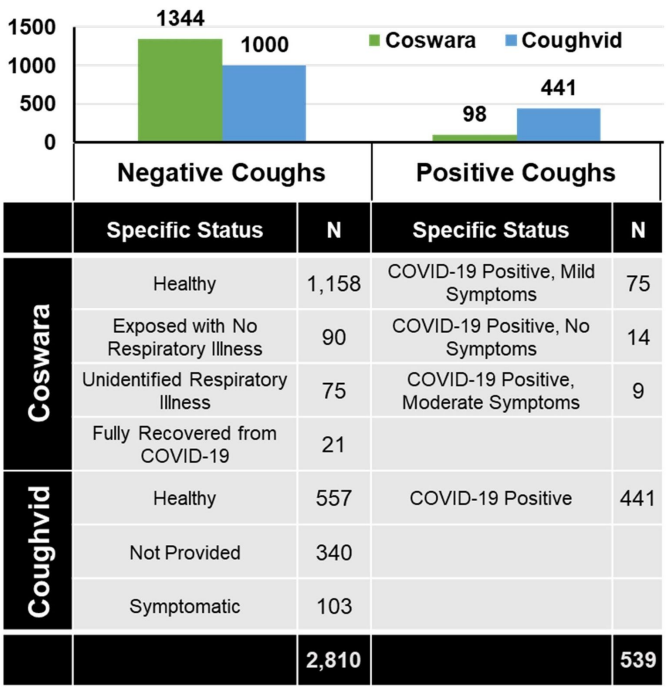

Fig. 2. Distribution of samples in the training dataset. PCR negative labels were not clearly indicated in either crowdsourced dataset. The Coswara dataset shows a heavy skew with mostly negative data.

### B. Virufy Test Datasets

To verify our model's performance beyond the Coswara and Coughvid crowdsourced data, we compiled two additional datasets with more detailed labels. All data had COVID-19 PCR labels and was acquired in conditions that were meant to simulate real-world usage. Audio files were a mixture of compressed and uncompressed files (e.g. wav and mp3 files) depending on the mode of data acquisition. We addressed potential privacy risks and security threats through localized privacy policies and patient consent forms, along with a Data Protection Impact Assessment (DPIA) and several internal information security policies.

*1) Virufy Latin American Crowdsourced Test Dataset*

To mimic one potential use case of COVID-19 detection from cough by smartphone users in the general public, samples used within the models were crowdsourced using the Virufy mobile data collection app (Fig. 1). For our analysis, we considered n=178 smartphone-recorded coughs from Peru, Brazil, and Colombia. Each cough sample was accompanied with specific labels of COVID-19 status based on PCR and antibody testing, demographics, past medical history, current symptoms, and a corresponding speech sample of counting from 1 to 10 (A.2). After excluding untested individuals and samples with poor audio quality, n=32 samples of PCR-negative and -positive individuals were aggregated to evaluate our algorithm's generalizability (Table. I).

*2) Virufy South Asian Clinical Test Datasets*

To determine performance of a COVID-19 detection algorithm in a busy clinical setting, we also collected samples in hospital clinics using smartphones. The explicit patient consent forms electronically accepted by all patients were originally drafted by Virufy clinical researchers and reviewed by our medical advisors. The data was captured directly from patients under the hospitals' Institutional Review Board (IRB) approved clinical research study protocols.

We collected Clinical dataset 1 from a South Asian hospital clinic from 04/13/2020 to 05/21/2020. It consists of cough recordings and labels from a total of 362 unique patients. Personal protective equipment and strict sanitation procedures were used to prevent disease transmission during this data acquisition. All patients were also simultaneously PCR tested for COVID-19 infection. Other demographics, medical history, and current symptoms information were also recorded (A.2).

As this data was acquired in a busy clinic, many samples had background noise, including clinician conversations, equipment sounds, ambient environmental noise, and vehicular traffic. In a manual survey of 50 random samples, 41 (82%) samples had at least one instance of distinctive noise.

We collected Clinical dataset 2 at a second hospital in a different country in South Asia from patients who had been PCR tested for COVID-19 with the same data points as in dataset 1. The data was collected from patients being screened at the fever clinic as well as from COVID-19 general and ICU wards.

These datasets were collected from a variety of smartphone types with different audio sampling frequencies, compression artifacts, and background noises.

### C. Audio and Clinical Features

We used multiple features from the crowdsourced datasets to train our network. After searching on various features and architectures using grid search, we found that an ensemble model of three features with the following parameters showed the best performance. The first feature was mel-frequency cepstral coefficients (MFCCs), a commonly used audio feature derived from the short-term power spectrum [17]. Each cough audio file was resampled to 22.5 kHz and the first 39 MFCCs were extracted using the librosa package [18], with a sampling rate of 22.5 kHz, hop length of 23ms, window length of 93ms, and a Hann window type. Outputs were averaged across the time-axis to yield mean 39 MFCCs features for each audio file.

The next extracted feature is the mel-frequency spectrogram, another common audio feature. Though MFCCs are derived from the spectrogram, the spectrogram encodes raw power information without any transformations. Spectrograms were extracted using the librosa package with the same parameters as for the MFCCs and interpolated to size (64,64).

Beyond audio files, each sample also contained additional rich information that has potential to enhance prediction accuracy. We chose to add two additional features for each cough file that reflect the clinical picture of the patient. Detectable changes in cough sounds have been shown to occur with diseases other than COVID-19 [19]. Therefore, a binary label about the presence or absence of current respiratory diseases was aggregated to feed into our algorithm as one extra feature. Next, COVID-19 also presents with other symptoms than cough, with some of the most common being fever and myalgia (muscle pain) [20]. The presence or absence of these symptoms may also impact the probability of having COVID-19. To develop as accurate a model as possible, a second binary label of fever or myalgia status was also aggregated from all datasets and fed into the model as a second extra feature. The distribution of each of these two extra features is shown in Fig. 3 below.

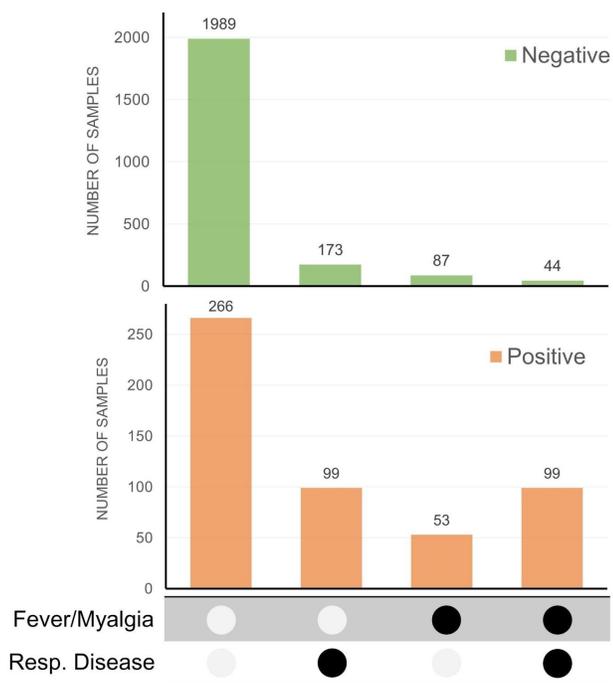

Fig. 3. Distribution of extra features in training set (n=3349). Fever/myalgia and medical history of respiratory disease were aggregated into two extra features. This figure shows the distributions for negative and positive samples separately for all patterns of these features.

### D. Ensembled Deep Neural Networks

After experimentation with 1D and 2D CNNs, LSTM, and CRNN architectures, the best performing network was an ensemble of 3 separate networks whose structure and hyperparameters were fine-tuned using grid search to minimize overfitting. Outputs from each network were aggregated to predict the probability of having COVID-19 (Fig. 4 below).

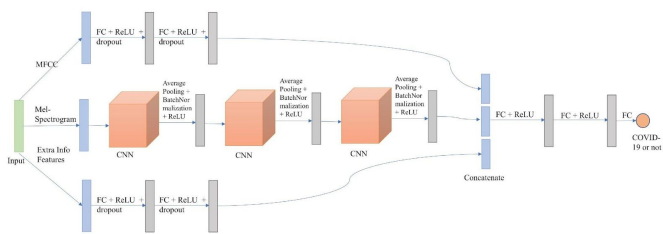

Fig. 4. Ensemble Model Structure.

The first network is for the MFCCs with input size of (39,) and consists of two hidden layers with ReLU activation, each followed by a dropout layer. The second network is a convolutional neural network with the mel-spectrogram image as input of size (64,64,1). It consists of three 2D convolution layers, with a kernel size of 3 and a stride size of 2 for the first convolution layer and a kernel size of 3 and a stride size of 1 for the rest two convolution layers, each followed by a 2D average pooling, a batch normalization, and a ReLU activation. The last network is for each sample's two extra features of fever/myalgia and respiratory conditions. Similar to the first network, it consists of two hidden layers with ReLU activations, each followed by a dropout layer. Outputs from each network were aggregated, fed through two additional hidden layers, each followed by a ReLU activation function, and combined into a final sigmoid output decision layer.

The ensemble network was trained using cross entropy loss, an Adam optimizer, and learning rate of 0.001. The training data was randomly split into train-validation-test datasets using a 70-15-15 split. Every experiment was repeated five times, each with a different random data split. The mean statistical values and 95% confidence intervals are reported, unless otherwise specified.

## IV. RESULTS

Table. II contains the test results on the four test datasets that are discussed above. We used both accuracy and Area under the ROC Curve (AUC) as evaluation metrics. As the data is unbalanced, we believe that AUC would be a better presentation of how the model is working. The model is robust enough that the test results on the Virufy Crowdsourced dataset, Clinical Dataset 1 and Clinical Dataset 2 are not significantly impacted by the change of the dataset. As the datasets are a mixture of compressed and uncompressed files and the compression downgrades the audio quality, a decrease in performance was expected. However, the Virufy Crowdsourced dataset and the Clinical Dataset 2 have both shown AUC results higher than 0.7, indicating that the model is generalizable to all four test datasets. The Clinical Dataset 1 demonstrates an AUC of 0.59, which correlates with the fact that the samples are noisy compared to the others but also shows promise in extremely noisy samples. Fig. 5 below illustrates the ROC curves of results on the four test datasets, which further confirms that our model generalizes to different datasets.

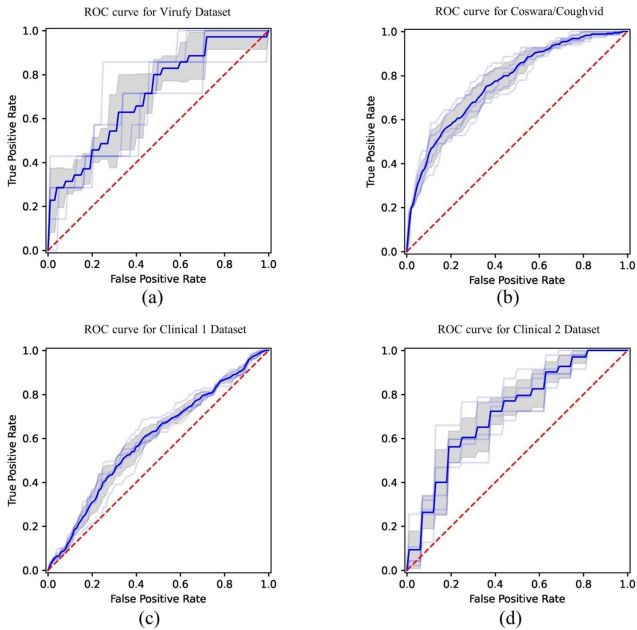

Fig. 5. ROC curves show high transfer learning of model to outside crowdsourced and clinical data. ROC curves of Virufy Crowdsourced Dataset, the Coswara/Coughvid subset, Clinical Dataset 1, and Clinical Dataset 2 are compared.

To prove our model is able to distinguish the difference between coughs from COVID-19 patients and normal coughs, we conducted a t-test on the AUC scores with the null hypothesis that the model can't differentiate between COVID-19 and non-COVID-19 recordings at a confidence level of 95%. Table. II shows all the P-Values calculated with the four test datasets and all of them show a result smaller than $\alpha = 0.05$. As exhibited, our tests have shown evidence that there is a significant statistical difference for all our datasets.

## V. CONCLUSIONS AND DISCUSSION

Using relevant audio features, the ensemble deep learning model was successful in identifying COVID-19 positive patients. We verified that a crowdsourced approach to collecting data can yield an accurate COVID-19 detection algorithm from cough. Since crowdsourced data is publicly available, we anticipate future work to build on our deep learning approaches and results.

We also demonstrate that our detection algorithm maintains its performance on external crowdsourced and clinical datasets which were collected using slightly different instructions in environments that were less than ideal and at different stages of infection. This demonstration gives credence to the hypothesis that COVID-19 can be reliably detected from cough sounds, as the virus signature appears to generalize.

Recording cough samples with no audio pollution in a clinical setting, especially in underserved countries, is not easily accomplished. In this study, we demonstrate that a cough-based COVID-19 detection algorithm can perform adequately well on noisy clinical data like our Clinical Dataset 2 (AUC=0.72), but has some limitations as demonstrated by its performance on Clinical Dataset 1 (AUC=0.59). We aim to continue technical work on improving performance on noisy data, with the hope that the eventual algorithm will work on cough data with often unavoidable background noise from hospital wards.

A limitation of this study is sample size: the datasets with detailed labels we selected for this study were not large enough to train the best performing algorithm or to conduct thorough subgroup and longitudinal analyses. To improve the performance of future machine learning algorithms, Virufy intends to collect larger quantities of high-quality cough data around the world to represent a wide range of ethnicities and community-specific phonological differences across populations. Most existing crowdsourced COVID-19 audio data collection efforts have been focused in Asia, Europe, and the United States. Although Latin America currently has among the highest rates of COVID-19 globally and accounts for 17.3% of current COVID-19 infections at the time of writing this manuscript [21], less than 5% of openly available cough samples are from Latin America [9,10]. We aim to continue our efforts for targeted data collection in Latin America to prevent the exclusion of this community from an ML-based solution.

However, most current approaches of data collection do not adequately account for variability, such as various settings, COVID-19 status labels, past medical history, and stage and severity of COVID disease. As COVID-19 has a highly variable presentation, including various combinations of anosmia, fever, asymptomatic low oxygen saturation, pneumonia, conjunctivitis, and heart injury [22,23], it is an open question whether any cough-based machine learning algorithm will be equally accurate for the entire spectrum of manifestations. However, a well-validated COVID-19 detection algorithm from cough has broad global applicability and can be instrumental in controlling the spread of the disease.

## VI. FUTURE WORK

A key challenge of clinical COVID-19 diagnosis is that its symptoms mimic those of other common respiratory, pulmonary, and cardiac conditions [24]. Therefore, further sub-analyses testing is necessary to determine the ability of machine learning algorithms to distinguish COVID-19 from other illnesses, such as non-COVID-19 pneumonia, upper/lower respiratory infections, asthma, and chronic lung disease exacerbations [19].

We are currently conducting longitudinal crowdsourced studies and clinical studies across various countries. Our goal is to train a machine learning algorithm with more information about human respiratory sound features, including cough and speech, both before symptom onset and over the course of COVID-19 infection. After gathering more audio data in association with PCR and evolving in vitro COVID-19 diagnostics, demographics, and disease course labels, we intend to conduct thorough sub-analyses that can validate an ML solution's performance in a multitude of conditions and demographic groups. As our current models are relatively shallow, we plan to develop deeper models as we collect more data from various contexts.

Although many groups worldwide are collecting various respiratory audio data with COVID-19 labels [9-15], at the time of this writing, most data is not openly available. To our knowledge, only Coswara [9] and Coughvid [10] have released open crowdsourced datasets, while only Virufy has released open clinical data. This confidentiality among groups working on the same technical challenge has hampered collaboration, reproducibility of results, and development of a widely available solution. Moving forward, broad collaboration and data sharing could promote the rapid development of AI-based COVID-19 detection tools.

To facilitate that effort, Virufy aims to establish a global consortium, bringing together research groups from around the world to share knowledge and datasets to build and rapidly refine AI algorithms to address the COVID-19 pandemic.

## APPENDIX

### A. Data Global distribution of training data (n=2748)

The crowdsourced datasets we used had cough samples from all around the world. Graph A below depicts this distribution for the subset of our training data that had location labels. The vast majority of Coswara dataset samples were from India, and most Coughvid dataset ones were from Europe and the United States. Notably, fewer samples were from Latin American and African countries. Fig. 6 and Fig. 7 below shows the number of samples from the top 20 countries.

### B. Detailed Characteristics of Test Datasets

Table. III contains several demographic and medical characteristics for the three test datasets used to show model generalizability. People may have several symptoms and medical conditions. (n.c. = not collected)

ACKNOWLEDGMENT

We are very grateful to Kara Meister, M.D., Stanford University Clinical Assistant Professor of Otolaryngology, and Mary L. Dunne, M.D., Stanford University Distinguished Career Institute Fellow, for their kind guidance on our clinical data collection procedures and analysis.

We are indebted to Dr. Jure Leskovec, Associate Professor of Computer Science at Stanford University, and Dr. Rok Sosic for their guidance on crowdsourcing and academic collaboration in the AI domain.

We appreciate Siddhi Hedge and Shreya Sriram for their amazing enthusiasm and hard work in facilitating clinical cough data collection from COVID-19 tested patients.

Furthermore, we thank the Coswara and Coughvid groups for open sourcing their COVID-19 datasets. We also thank everyone who has contributed their cough data to Virufy.

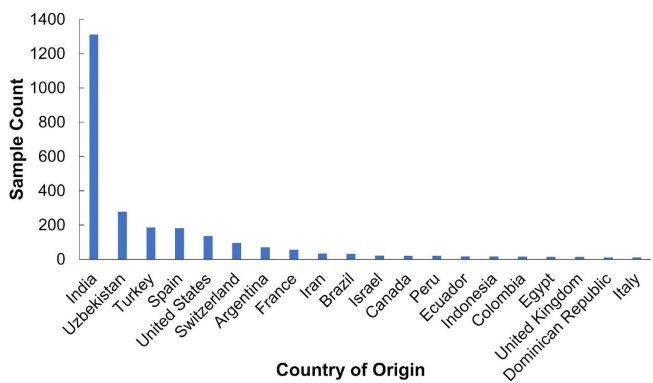

Fig. 5. Country of origins distributions.

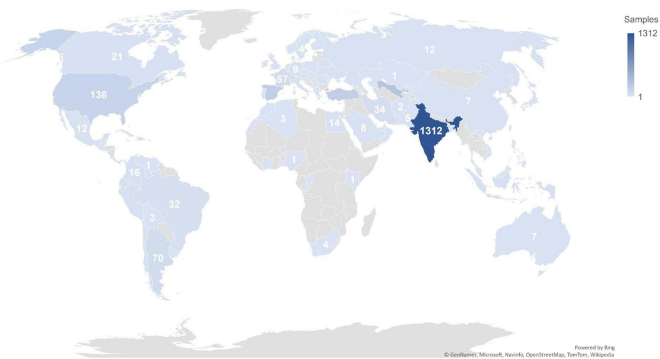

Fig. 6. Distribution Map.

TABLE I
TABLE FOR TEST DATA DISTRIBUTION

| PCR Result | Virufy Crowdsourced | Clinical Dataset 1 | Clinical Dataset 2 |
|---|---|---|---|
| Positive | 7 (22.6%) | 89 (24.6%) | 47 (74.6%) |
| Negative | 24 (77.4%) | 273 (75.4%) | 16 (25.4%) |
| Total | 31 | 362 | 63 |

Composition of label distribution and counts of Test Datasets

TABLE II
TABLE FOR EXPERIMENT RESULTS

| | Coswara /Coughvid | Virufy Crowd sourced | Virufy Clinical Dataset 1 | Virufy Clinical Dataset 2 |
|---|---|---|---|---|
| AUC | 0.771 | 0.721 | 0.586 | 0.718 |
| P-Value | 0.001 | 0.002 | 0.018 | 0.0003 |
| CI | 0.739- 0.802 | 0.678- 0.764 | 0.614- 0.768 | 0.674- 0.763 |

Experiment Results, including AUC scores, P-Values, Confidence Intervals (CI) on various datasets: Coswara/Coughvid subset, Virufy Crowdsourced, Clinical Dataset 1 and Clinical Dataset 2.

TABLE III
TABLE FOR DEMOGRAPHIC AND MEDICAL DISTRIBUTION

| | Virufy Crowdsourced (n=31) | Clinical Dataset 1 (n=362) | Clinical Dataset 2 (n=63) |
|---|---|---|---|
| **PCR Test Result (%)** | | | |
| Positive | 7 (22.6) | 89 (24.6) | 47 (74.6) |
| Negative | 24 (77.4) | 273 (75.4) | 16 (25.4) |
| **Sex (%)** | | | |
| Female | 11 (35.4) | 117 (32.3) | 22 (34.9) |
| Male | 19 (61.2) | 244 (67.4) | 41 (65.1) |
| Other | 1 (0.03) | 1 (0.3) | 0 (0) |
| **Age (%)** | | | |
| 0-9 | 0 (0) | 1 (0.3) | 0 (0) |
| 10-19 | 2 (6.5) | 15 (4.1) | 5 (7.9) |
| 20-29 | 15 (48.4) | 123 (34) | 30 (47.6) |
| 30-39 | 4 (12.9) | 86 (23.8) | 12 (19) |
| 40-49 | 2 (6.5) | 62 (17.1) | 5 (7.9) |
| 50-59 | 5 (16.1) | 56 (15.5) | 6 (9.5) |
| 60-69 | 2 (6.5) | 18 (5) | 4 (6.3) |
| 70-79 | 1 (3.2) | 1 (0.3) | 1 (1.6) |
| **Smoker (%)** | | | |
| Yes | 6 (19.4) | 69 (19.1) | 17 (27) |
| No | 25 (80.6) | 293 (80.9) | 46 (73) |
| **Symptoms (%)** | | | |
| None | 20 (64.5) | 188 (51.9) | 11 (17.5) |
| New or worsening cough | 3 (9.7) | 92 (25.4) | 24 (38.1) |
| Sore throat | 3 (9.7) | 84 (23.2) | 28 (44.4) |
| Fever and/or chills | 2 (6.5) | 83 (22.9) | 40 (63.5) |
| Dyspnea | 3 (9.7) | 73 (20.2) | 2 (3.2) |
| Anosmia | 1 (3.2) | 38 (10.5) | 5 (7.9) |
| Myalgia | 3 (9.7) | 22 (6.1) | 27 (42.9) |
| Emesis and/or diarrhea | 1 (3.2) | 4 (1.1) | 4 (6.3) |
| Headaches | 5 (16.1) | n.c. | 22 (34.9) |
| **Medical Conditions (%)** | | | |
| None | 27 (87.1) | 302 (83.4) | 57 (90.5) |
| Asthma | 3 (9.7) | 18 (5) | 1 (1.6) |
| Chronic Tuberculosis | 1 (3.2) | 0 (0) | 0 (0) |
| Thrombophilia | 1 (3.2) | 0 (0) | 0 (0) |
| Extreme Obesity | 1 (3.2) | 0 (0) | 1 (1.6) |
| Congestive Heart Failure | 0 (0) | 12 (3.3) | 0 (0) |
| Diabetes | 0 (0) | 31 (8.6) | 5 (7.9) |
| Hypertension | 0 (0) | n.c. | 8 (12.7) |